\documentclass[usenatbib]{mn2e}
\usepackage{apjfonts,amsfonts,amsmath,amssymb,bm,ctable,verbatim}

\newcommand{\Alf}{{Alfv\'en}}

\newcommand{\bhat}{\hat{\bf b}}

\newcommand{\etal}{et al.}

\newcommand{\acknowledgments}[1]{\begin{small}\section*{Acknowledgments}\end{small}{\noindent #1}\vspace{5pt}}
\newcommand{\datastatement}[1]{\begin{small}\section*{Data Availability Statement}\end{small}{\noindent #1}\vspace{5pt}}

\title[Improving Piecewise-Power-Law Diffusion]{An accurate treatment of scattering and diffusion in piecewise power-law models for cosmic ray and radiation/neutrino transport}

\author[Hopkins \etal]{
\parbox[t]{\textwidth}{
Philip F.~Hopkins$^1$, 
}\vspace*{4pt} \\
$^1$ TAPIR, Mailcode 350-17, California Institute of Technology, Pasadena, CA 91125, USA. E-mail:phopkins@caltech.edu \\
}

\date{}
\begin{document}
\maketitle

\begin{abstract}
A popular numerical method to model the dynamics of a ``full spectrum'' of cosmic rays (CRs), also applicable to radiation/neutrino hydrodynamics (RHD), is to discretize the spectrum at each location/cell as a piecewise power law in ``bins'' of momentum (or frequency) space. This gives rise to a pair of conserved quantities (e.g.\ CR number and energy) which are exchanged between cells or bins, that in turn give the update to the normalization and slope of the spectrum in each bin. While these methods can be evolved exactly in momentum-space (e.g.\ considering injection, absorption, continuous losses/gains), numerical challenges arise dealing with {\em spatial} fluxes, if the scattering rates depend on momentum. This has often been treated by either by neglecting variation of those rates ``within the bin,'' or sacrificing conservation -- introducing significant errors. Here, we derive a rigorous treatment of these terms, and show that the variation within the bin can be accounted for accurately with a simple set of scalar correction coefficients that can be written entirely in terms of other, explicitly-evolved ``bin-integrated'' quantities. This eliminates the relevant errors without added computational cost, has no effect on the numerical stability of the method, and retains manifest conservation. We derive correction terms both for methods which explicitly integrate flux variables (e.g.\ two-moment or M1-like) methods, as well as single-moment (advection-diffusion, FLD-like) methods, and approximate corrections valid in various limits. 
\end{abstract}


\section{Introduction}
\label{sec:intro}

Understanding cosmic ray (CR) propagation and dynamics in the interstellar medium (ISM) and circum/inter-galactic medium (CGM/IGM) remains an unsolved problem of central importance in space plasma physics \citep{Zwei13,zweibel:cr.feedback.review,2018AdSpR..62.2731A,2019PrPNP.10903710K}, with major implications for fields ranging from astro-chemistry, planet, star, and galaxy formation \citep[e.g.][]{Chen16,Simp16,Giri16,Pakm16,Sale16,wiener:2017.cr.streaming.winds,Rusz17,Buts18,farber:decoupled.crs.in.neutral.gas,Jaco18,chan:2018.cosmicray.fire.gammaray,Buts18,su:turb.crs.quench,hopkins:cr.mhd.fire2,ji:fire.cr.cgm,ji:20.virial.shocks.suppressed.cr.dominated.halos,bustard:2020.crs.multiphase.ism.accel.confinement}. 

In models which seek to {\em dynamically} evolve the CR population on large scales (as opposed to either historical semi-analytic models, which solve for the equilibrium CR distribution function (DF) in a static analytic Galaxy model, e.g.\ \citealt{2016PhRvD..94l3019K,evoli:dragon2.cr.prop,2018ApJ...869..176L,2018AdSpR..62.2731A}, or particle-in-cell type simulations which model the dynamics of individual CRs), a central challenge is the high dimensionality of the DF $f({\bf x},\,{\bf p},\,t)$ as a function of position ${\bf x}$, CR momentum ${\bf p}$, and time $t$. Recently, a number of studies \citep{ogrodnik:2021.spectral.cr.electron.code,hanasz:2021.cr.propagation.sims.review,hopkins:cr.multibin.mw.comparison,girichidis:2021.cr.transport.w.spectral.reconnection.hack} have addressed this by implementing variations of the method proposed in \citet{girichidis:cr.spectral.scheme} (with broadly similar methods used earlier in e.g.\ \citealt{1999ApJ...511..774J,miniati:2001.cr.piecewise.powerlaw,2007JCoPh.227..776M,2001ApJ...562..233M,jones:2005.cr.shock.spectral.sims,mimica:2009.spectral.evol.crs.jets,yang:2017.fermi.cr.spectra.sims,winner:2019.cr.electron.post-processing.tracers} as well), wherein the isotropic part of the DF $\bar{f}_{0}$ is represented as a piecewise power-law function of momentum, in ``bins'' of $p$ spanning some dynamic range; one can then integrate (to arbitrary precision) bin-to-bin fluxes of conserved CR number and energy (representing e.g.\ continuous loss or gain processes) or source/sink terms (injection or catastrophic losses or secondary production) in momentum space. 

The method has many advantages. (1) Because real CR spectra are smooth and power-law-like over a wide dynamic range, these studies have shown that the spectrum over some very wide dynamic range can be represented accurately with a relatively small number of bins per species, imposing modest computational and memory cost. (2) The momentum-space and coordinate-space (advection/streaming/diffusion) operations can be operator-split, allowing the spatial part of the equations to be integrated with standard, well-studied and high-order numerical methods (exactly identical to previous treatments which considered just a single CR ``fluid'' or bin or total energy density scalar field, e.g.\ \citealt{Sale16,Rusz17,chan:2018.cosmicray.fire.gammaray,Buts18,su:turb.crs.quench,hopkins:cr.mhd.fire2,ji:fire.cr.cgm,ji:20.virial.shocks.suppressed.cr.dominated.halos,bustard:2020.crs.multiphase.ism.accel.confinement}). (3) Conservation of number and energy is manifest, which ensures robustness of many results even in highly noisy conditions or in extreme injection/loss events. (4) It is accurate and converges efficiently in momentum-space. (5) It trivially generalizes for methods which evolve either the  ``two-moment'' equations  for the CR DF (where one evolves both the isotropic part of the DF and its flux, or equivalently the mean CR pitch angle), or ``one-moment'' equations (where one assumes the flux is in local steady-state, so evolve just the isotropic part of the DF subject to a diffusion+streaming equation), as well as to even-further-simplified models (e.g.\ replacing the correct anisotropic diffusion with isotropic diffusion). These and other advantages have led, for example, to the first simulations simultaneously evolving multi-species CR spectra alongside ``live'' fully-coupled MHD dynamics on Galactic scales \citep{hopkins:cr.multibin.mw.comparison}. 

However, while the {\em momentum}-space properties of this class of piecewise-power-law methods are very well-defined (and easy to demonstrate), there is a known conceptual challenge in {\em coordinate}-space. Specifically, given some piecewise-power-law representation of $f$ in a ``bin,'' the spatial flux of $f$ should depend on momentum, varying ``across the bin.'' But since the flux depends itself on gradients of various moments of the distribution function itself, a naive attempt to integrate or average the flux over the bin leads to expressions of the form $\int\,d^{3}\,{\bf p}\,\mathbb{F}[{\bf p},\,f[{\bf p}],\,....]\,\cdot \nabla \cdot \mathbb{G}[{\bf p},\,f[{\bf p}],\,...]$, where $\mathbb{F}$, $\mathbb{G}$ are some arbitrary tensor functions. These are not just complicated, but appear at first to require ``sub-binning'' of $f$ into infinitesimally small bins, each of which has a separately-computed gradient, in order to evaluate accurately \citep{girichidis:cr.spectral.scheme}. As a result, most studies above have adopted the ``bin-centered'' approach, wherein one assumes that all quantities of relevance for computing spatial fluxes are assumed to be constant over the momentum-width of a bin. This retains advantages (1), (2), (3), and (5) above, but leads to well-known artifacts in the spectrum when spatial transport (e.g.\ diffusion) dominates the escape time, sacrificing some of (4). Alternative approaches have been discussed (e.g.\ \citealt{girichidis:2021.cr.transport.w.spectral.reconnection.hack}), but (as noted by these authors) these generally sacrifice all of (2), (3), and (5); in particular the proposed non-bin-centered methods sacrifice conservation and consistency (they cannot be derived from the underlying DF equations) and can potentially lead to numerical instability or unphysical behaviors when momentum-space terms (e.g.\ losses) dominate. 

In this paper we derive a consistent treatment of these terms which resolves all of the challenges above and retains all of advantages (1-5) above. By considering a two-moment pitch-angle expansion of the Vlasov equation on scales large compared to CR gyro-radii, we show that the key conceptual ingredient required to resolve these issues is a consistent treatment of how the mean CR pitch angle varies across a ``bin.'' But we also show that the structure of the equations imposes consistency conditions which determine this at the level of approximation needed for the piecewise-power-law reconstruction. With this properly treated, we show the corrected numerical method is structurally identical to the ``bin-centered'' approximation with appropriate scalar correction coefficients which are determined entirely in terms of already-evolved numerical quantities. We further show that the correction coefficients can be (self-consistently) even-further simplified if either (1) only the one-moment equation for the CRs is dynamically evolved, or (2) one only needs to capture the exact behavior in all relevant limits of the local-steady-state flux equation (e.g.\ one is interested primarily in timescales long compared to CR scattering times). 

While our primary motivation in this paper is focused on applications to CRs, this qualitative method, and the challenges above, also apply in principle to analogous methods which evolve spectra of other collisionless species (e.g.\ radiation or neutrinos) as piecewise-power-laws in similar fashion (e.g.\ \citealt{1997A&A...320..920B}). In this context, most ``moment-based'' multi-group methods for radiation-hydrodynamics (RHD) have focused on evolving just the radiation/neutrino energy in each ``bin'' \citep[e.g.][]{castor:2007.rhd.book}, effectively equivalent to representing the spectrum as piecewise-constant, rather than a piecewise power-law. Although conceptually simpler, the piecewise-constant approach requires an order of magnitude larger number of ``bins'' across some frequency or energy range in order to represent spectra with steep or dynamically-evolving power-law slopes, and sacrifices the ability to simultaneously conserve number and energy. A method like the piecewise-power-law scheme above for neutrinos has been discussed in e.g.\ \citet{rampp:2002.neutrino.rhd.with.number.egy.con,muller:2010.multigroup.rhd.number.egy.cons.problems} (their ``simultaneously number-and-energy-conserving scheme,'' although it is described in different language than we use here), but similar conceptual difficulties (see \citealt{mezzacappa:2020.neutrino.transport.review}) have limited its application.

\section{A Method for Handling Fluxes of Piecewise-Power-Law Spectra}

\subsection{Setup \&\ Definitions}
\label{sec:setup}

Consider a population of CRs\footnote{For our purposes here, different species of CR are linearly independent so it is sufficient to consider the DF for a single species (the total DF can then be reconstructed by simply summing over species).} with some phase-space distribution function (DF) $f = dN_{\rm cr}/d^{3}{\bf x}\, d^{3}{\bf p}$, with polar momentum coordinates $p=|{\bf p}|$, pitch angle $\mu \equiv \cos{\theta} \equiv \hat{\bf p} \cdot \bhat$ (where $\bhat \equiv {\bf B}/|{\bf B}|$ is the magnetic field direction), and phase angle $\phi_{\rm g}$. The comoving evolution equations for the {\em spatial} or coordinate-space part of the first two $\mu$-moments of $f$ can be written \citep{hopkins:m1.cr.closure}:\footnote{Eq.~\ref{eqn:spatial} formally follows from the Vlasov equation, with the standard quasi-linear scattering terms from \citet{schlickeiser:89.cr.transport.scattering.eqns}, assuming the DF is approximately gyrotropic, expanding to leading order in  $\mathcal{O}(r_{\rm gyro}/L_{\rm macro})$ (the ratio of gyro radius to resolved macroscopic scales) and $\mathcal{O}(|{\bf v}_{\rm gas}|/c)$ (ratio of background MHD bulk velocities to $c$).}
\begin{align}
\label{eqn:spatial} D_{t} \bar{f}_{0} &= -\nabla \cdot (v\,\bar{f}_{1}\,\bhat) + ....\  , \\
\label{eqn:spatial.flux} D_{t} \bar{f}_{1} &+ v\,\bhat \cdot \nabla \cdot (\mathbb{D}\,\bar{f}_{0}) = -  \bar{D}_{\mu\mu}\,\bar{f}_{1} - \bar{D}_{\mu p}\,{\partial_{p} \bar{f}_{0}} + ...\ .
\end{align} 
where $\bar{f}_{n} \equiv (4\pi)^{-1}\,\int d\mu\,d\phi_{\rm g}\,\mu^{n}\,f$, so $\bar{f}_{0}$ is the isotropic part of the DF and $\bar{f}_{1} \equiv \langle \mu \rangle\,\bar{f}_{0}$; $D_{t} X \equiv \partial_{t} X + \nabla \cdot ({\bf v}_{\rm gas}\,X) = \rho\,d_{t}(X/\rho)$; $\chi \equiv (1 - \langle \mu^{2} \rangle)/2$ and $\mathbb{D} \equiv \chi\,\mathbb{I} + (1-3\,\chi)\,\bhat\otimes\bhat$; $\bar{D}_{\mu\mu} \equiv \bar{\nu}$ is the pitch-angle averaged scattering rate (at the given $p$ and ${\bf x}$); and $\bar{D}_{\mu p} \equiv \bar{\nu}\,\chi\,p\,\bar{v}_{A}/v$ in terms of the CR velocity $v = \beta\,c$ and $\bar{v}_{A} \equiv v_{A}\,(\bar{\nu}_{+} - \bar{\nu}_{-})/(\bar{\nu}_{+}+\bar{\nu}_{-})$ in terms of the ``forward'' and ``backward'' scattering coefficients $\nu_{\pm}$ and phase speed $v_{A}$ of gyro-resonant \Alf\ waves (those with wavelength $\sim r_{\rm gyro}$). We stress that Eqs.~\ref{eqn:spatial}-\ref{eqn:spatial.flux} are valid for any arbitrary gyrotropic DF: different ``closure'' assumptions relate to how $\langle \mu^{2} \rangle$ is specified \citep[see][]{hopkins:m1.cr.closure}, which is not important for our purposes.

In Eqs.~\ref{eqn:spatial}-\ref{eqn:spatial.flux}, the ``...'' refers to terms which do not propagate CRs in coordinate space (e.g.\ injection \&\ catastrophic losses $D_{t} f = j$, and continuous energy loss/gain processes $D_{t} f = p^{-2}\,\partial_{p}[p^{2}\,...]$). These can be operator-split and solved accurately with methods like those in \S~\ref{sec:intro} \citep{girichidis:cr.spectral.scheme,ogrodnik:2021.spectral.cr.electron.code,hanasz:2021.cr.propagation.sims.review,hopkins:cr.multibin.mw.comparison,girichidis:2021.cr.transport.w.spectral.reconnection.hack}, which model the spectrum as a piecewise-power-law. In these methods, within some infinitesimally small volume domain $j$, for each CR species $s$, within some ``bin'' $m$ defined over a momentum interval $p^{-} < p < p^{+}$, we assume that $\bar{f}_{0}$ can be represented as a power-law with slope $\alpha_{f_{0}}$, i.e.: 
\begin{align}
\label{eqn:f0.power.law} \bar{f}_{0,\,j,m,s} &\approx (\bar{f}_{0})_{0,\,j,m,s}\,\left( \frac{p}{p_{0,\,m,s}} \right)^{\alpha_{f_{0},\,j,m,s}} 
\end{align}
where for analytic convenience we define $p_{0} \equiv (p^{+}\,p^{-})^{1/2}$ as the geometric mean momentum of the ``bin.'' 
It is immediately obvious that the spatial part of Eqs.~\ref{eqn:spatial}-\ref{eqn:spatial.flux} is independent for each ``bin'' $m$ and species $s$ (i.e.\ there is no cross-term in Eqs.~\ref{eqn:spatial}-\ref{eqn:spatial.flux} coupling different species or momenta), so we only need to consider one such bin to completely specify the numerical method. We therefore drop the $j,m,s$ notation for brevity, with the understanding that all quantities considered here can (and should) depend on $s$, $m$, and spatial location. 

For reference below we also define $\xi\equiv p^{+}/p^{-}$ as a dimensionless ``bin width.''

\subsection{Conserved Quantities and the Spatial Flux}

Given our power-law representation of $\bar{f}_{0}$ in Eq.~\ref{eqn:f0.power.law} with two parameters ($(\bar{f}_{0})_{0}$ and $\alpha_{f_{0}}$), we can clearly represent or evolve exactly two independent conserved scalar quantities of the DF (and their associated fluxes as we show below) associated with each bin. These are typically chosen to be the CR number and (kinetic) energy, with volumetric densities $n$, $\epsilon$.\footnote{We can freely choose to evolve the kinetic or total CR energy, since given the CR number they are trivially related. Here and in most applications the kinetic energy is preferable because in the non-relativistic limit, determining the kinetic energy via subtracting the rest energy from the total energy (two large numbers) can lead to fractionally large floating-point errors.} We can define the density of any such scalar quantity in the bin by: 
\begin{align}
q &\equiv \int d^{3}{\bf p}\,\phi_{q}\,f = \int_{p^{-}}^{p^{+}}\,4\pi\,p^{2}\,dp\,\phi_{q}\,\bar{f}_{0}
\end{align}
where for $q=(n,\,\epsilon)$ we have $\phi_{q} = (1,\,T[p])$ (with $T \equiv (p^{2}\,c^{2} + m_{s}^{2}\,c^{4})^{1/2} - m_{s}\,c^{2}$ for rest mass $m_{s}$). 
So evolving ($(\bar{f}_{0})_{0}$, $\alpha_{f_{0}}$) is equivalent to evolving $(n,\,\epsilon)$. 
Returning to Eq.~\ref{eqn:spatial}, multiplying by $4\pi\,p^{2}\,dp\,\phi_{q}$ and integrating we immediately have: 
\begin{align}
\label{eqn:Dtq} D_{t} q &= - \nabla \cdot {\bf F}_{q} + .... \\ 
{\bf F}_{q} &\equiv \bhat\,F_{q} = \bhat\,\int_{p^{-}}^{p^{+}}\,4\pi\,p^{2}\,dp\,v\,\phi_{q}\,\bar{f}_{1}
\end{align}
which is a standard hyperbolic conservation equation that can be integrated to desired accuracy, provided an expression for $F_{q}$.\footnote{We can trivially turn Eq.~\ref{eqn:Dtq} into a flux equation for the volume-integrated conserved quantities of CR number or energy ($Q_{j} = (N_{j},\,\mathcal{E}_{j}) =  \int_{V_{j}} d^{3}{\bf x}\,q$) by integrating over some volumetric domain $V_{j}$ in usual finite-volume fashion, giving ${\rm d}_{t} Q_{j} = -\oint_{\partial j} {\bf F}_{q} \cdot {\bf A}$.} 

Conversely, since the DF in Eq.~\ref{eqn:f0.power.law} has two parameters which vary in space and time: $(\bar{f}_{0})_{0}$ and $\alpha_{f_{0}}$, in order to update both in a timestep self-consistently in a manifestly conservative manner, we must update both $(q,\,q^{\prime})=(n,\,\epsilon)$, which requires computing both fluxes ($F_{q}$, $F_{q^{\prime}}$). The updated $(n,\,\epsilon)$ in some next timestep then immediately give the new ($(\bar{f}_{0})_{0}$, $\alpha_{f_{0}}$). For details, see \citet{girichidis:cr.spectral.scheme}.

In principle, any ``basis function'' representation of $f(p)$ in the bin with two free parameters (of which a power-law is simply most convenient, given the real shape of the CR DF) should allow us to conserve two scalar quantities (CR number, energy) from evolving Eq.~\ref{eqn:Dtq}. If we also explicitly evolve the corresponding flux equations $D_{t} F_{q}$ (derived below), then we should also conserve both of their fluxes (i.e.\ the CR number and energy flux, which correspond to the CR current and momentum density fields).

\subsection{The Flux Evolution Equation}

So, taking Eq.~\ref{eqn:spatial.flux}, multiplying by $4\pi\,p^{2}\,dp\,v\,\phi_{q}$ and integrating, we have for the flux equation:
\begin{align}
D_{t} F_{q} &+ \bhat \cdot \nabla \cdot \mathcal{I}_{\nabla,q} = -\mathcal{I}_{0,q} -\mathcal{I}_{1,q} \\ 
\mathcal{I}_{\nabla,q} &\equiv \int_{p^{-}}^{p^{+}}\,4\pi\,p^{2}\,dp\,v^{2}\,\phi_{q}\,\mathbb{D}\,\bar{f}_{0} \\ 
\mathcal{I}_{0,q} &\equiv \int_{p^{-}}^{p^{+}}\,4\pi\,p^{2}\,dp\,v\,\phi_{q}\,\bar{D}_{\mu p}\,\frac{\partial \bar{f}_{0}}{\partial p} \\ 
\nonumber &= \int_{p^{-}}^{p^{+}}\,4\pi\,p^{2}\,dp\,\bar{\nu}\,\chi\,\alpha_{f_{0}}\,\bar{v}_{A}\,\phi_{q}\,\bar{f}_{0} \\ 
\mathcal{I}_{1,q} &\equiv \int_{p^{-}}^{p^{+}}\,4\pi\,p^{2}\,dp\, \phi_{q}\,\bar{D}_{\mu\mu}\,v\,\bar{f}_{1}  \\ 
\nonumber &= \int_{p^{-}}^{p^{+}}\,4\pi\,p^{2}\,dp\,\bar{\nu}\,v\,\langle \mu \rangle\,\phi_{q}\,\bar{f}_{0}
\end{align}
where we made use of various definitions above.
Now define, for any quantity $X$ which might vary as a function of $p$, $X_{0} \equiv X[p=p_{0}]$ (i.e.\ $X_{0}$ is the value of $X$ at the bin center). 
We can then immediately {define} the integral $\mathcal{I}$ terms in the following convenient form:
\begin{align}
\label{eqn:Igrad.omega} \mathcal{I}_{\nabla,q} &\equiv \omega_{\nabla,q}\,v_{0}^{2}\,\mathbb{D}_{0}\,q \\ 
\mathcal{I}_{0,q} &\equiv \omega_{0,q}\,\bar{\nu}_{0}\,\chi_{0}\,\alpha_{f_{0}}\,\bar{v}_{A,\,0}\,q  \\ 
\mathcal{I}_{1,q} &\equiv \omega_{1,q}\,\bar{\nu}_{0}\,F_{q}
\end{align}
which places the complicated integrals into the dimensionless functions $\omega$ (define by the above relations to $\mathcal{I}$).
This allows us to write the flux equation in familiar form:
\begin{align}
\label{eqn:DtF.familiar} D_{t} F_{q} + {v}_{0}^{2}\,\bhat \cdot \nabla \cdot \left( {\mathbb{D}}^{\rm eff}_{0,q}\,q \right) 
&= -{\nu}^{\rm eff}_{0,q}\,\left( F_{q} - {v}^{\rm eff}_{{\rm st},q}\,q\right) 
\end{align}
with the modified ``effective'' coefficients:
\begin{align}
{\mathbb{D}}^{\rm eff}_{0,q} &\equiv   \omega_{\nabla,q}\,\mathbb{D}_{0} \\ 
{\nu}^{\rm eff}_{0,q} &\equiv  \bar{\nu}_{0}\,\omega_{1,q} \\ 
{v}^{\rm eff}_{{\rm st},q} &\equiv -\frac{\omega_{0,q}}{\omega_{1,q}}\,\chi_{0}\,\alpha_{f_{0}}\,\bar{v}_{A,0}
\end{align}

\subsection{The Bin-Centered Approximation}
\label{sec:bin.centered}

As discussed in \S~\ref{sec:intro}), Eq.~\ref{eqn:DtF.familiar} has largely been evolved according to the ``bin-centered'' approximation, which evaluates $F_{q}$ as if we had an infinitesimally narrow bin centered at $p=p_{0}$, i.e.\ taking $\omega_{\nabla,q}=\omega_{1,q}=\omega_{0,q}=1$. This has obvious advantages: (1) it is numerically straightforward: in fact the spatial (advection+flux) equations for a single CR ``bin'' become numerically exactly identical to the ``single-bin'' CR equations (wherein one integrates over the entire CR spectrum and simply evolves a ``total CR energy''); (2) it is fairly trivially stable and robust (any integration method which can handle the two-moment equations for single-bin CRs, or radiation, or the one-moment diffusion+streaming equation, is trivially numerically stable and robust here); (3) it is simple; (4) it still retains manifest conservation: one still evolves both $F_{q}$ and $F_{q^{\prime}}$ (so e.g.\ can manifestly conserve CR number and energy as desired), with $F_{q}/q = F_{q^{\prime}}/q^{\prime}$ ($\Psi_{qq^{\prime}}\rightarrow 1$,  as defined below) required for consistency in this approximation (since we have taken the limit $|p^{+}-p^{-}|/|p^{+}+p^{-}| \rightarrow 0$ or $\ln{\xi}=\ln{(p^{+}/p^{-})} \rightarrow 0$, or $v\,\langle\mu\rangle=$\,constant across the bin, by definition). 

The problem with this approximation is that is not consistent with a non-trivial variation of $\langle \mu \rangle$ as a function of $p$ {\em within the bin}. Specifically, from the above, this assumes the CR drift velocity ($v\,\langle\mu\rangle$) is constant over each bin width. As a result, a piecewise power-law spectrum at injection (ignoring losses or any other effects besides pure spatial flux) will advect conserving the local power-law slope $\alpha_{f_{0}}$ in each bin. But if $\bar{\nu}$ is a decreasing function of $p$ (as physically expected), the advection speed of higher-$p$ bins will be faster than lower-$p$ bins, so (for a fixed injection rate) their equilibrium abundance will be lower, steepening the spectrum bin-to-bin. But since the slopes within each bin are conserved by flux in this approximation, one ends up with a spectrum that features a series of ``step''-like features between each bin \citep[see e.g.][]{girichidis:cr.spectral.scheme,girichidis:2021.cr.transport.w.spectral.reconnection.hack,ogrodnik:2021.spectral.cr.electron.code,hopkins:cr.multibin.mw.comparison}. We stress that these errors are usually small, and only apply when diffusive transport is the fastest loss/escape timescape (other loss/gain terms in these methods do modify the CR slopes, and as we show below, in the CR streaming limit, the correct behavior actually is equivalent to the bin-centered approximation). Effectively, in flux-steady-state (see \S~\ref{sec:local.flux.steady} below) in the highly-relativistic limit (the case of greatest interest), the bin-centered fluxes are formally what we would obtain if the diffusivity $\nu$ were a piecewise-constant function of $p$ (constant across each bin). But of course, that is not usually the desired model.

Because as we will show below, all of the correction terms $\omega \approx 1 + (...)\,|\ln{\xi}|^{2} + ...$ deviate from unity at $\mathcal{O}(|\ln{\xi}|^{2})$, the error here is formally second-order in momentum-space and would converge to some desired accuracy if we simply increased the number of bins to make $|\ln{\xi}|$ sufficiently small. But in most applications, that is computationally prohibitive.

\subsection{Towards a Better Approximation}

To do better, we must evaluate the correction terms $\omega$ for finite $\ln{\xi}$. By definition, most of the necessary inputs ($\phi_{q}$, $\bar{f}_{0}$, $v$) and their dependence on $p$ are specified. However the challenge is that all three $\omega$ terms depend on powers of $\mu$ (through $\bar{f}_{1}$ or $\chi$, $\mathbb{D}_{0}$). This introduces new variables whose dependence on $p$ (via $\mu$) is not {\em a priori} specified. 

\subsubsection{Terms which Depend Weakly on Pitch Angle}

Let us begin with $\omega_{\nabla}$. This depends only on specified inputs as above and $\mathbb{D}_{0}$, which depends on $\langle \mu^{2} \rangle$ through $\chi$. But here we can make use of the limiting behaviors of $\mathbb{D}$: for DFs which are near isotropic (hence $\langle \mu \rangle \rightarrow 0$ is small), $\chi \rightarrow 1/3 + \mathcal{O}(\langle \mu \rangle^{2})$ so $\mathbb{D} \rightarrow \mathbb{I}/3 + \mathcal{O}(\langle \mu\rangle^{2})$, while for DFs which are near maximally anisotropic/coherently free-streaming from a source ($\langle \mu \rangle \rightarrow \pm 1$), $\chi \rightarrow \mathcal{O}([|\langle \mu \rangle|-1]^{2})$ so $\mathbb{D} \rightarrow \bhat\otimes\bhat + \mathcal{O}([|\langle \mu \rangle|-1]^{2})$. In either regime, the dependence on $\langle \mu \rangle$ is quite weak, so even if $\langle \mu \rangle$ varies across the bin, it will produce very little variation in $\mathbb{D}$. So long as we do not see a very rapid transition from confinement to free-streaming across a single bin (which we do not expect), then it is almost always safe to neglect the variation in $\chi$ and $\mathbb{D}$ across any reasonable spectral bin size, i.e.\ take $\mathcal{I}_{\nabla,q} \approx \mathbb{D}_{0}\,\int_{p^{-}}^{p^{+}}\,4\pi\,p^{2}\,dp\,v^{2}\,\phi_{q}\,\bar{f}_{0}$. If we do this, then Eq.~\ref{eqn:Igrad.omega} immediately yields:
\begin{align}
\omega_{\nabla,q} &\approx \frac{\int_{{p}^{-}}^{{p}^{+}}\,d{p}\,{p}^{2}\,\bar{f}_{0}\,{v}^{2}\,{\phi}_{q}}{\int_{{p}^{-}}^{{p}^{+}}\,d{p}\,{p}^{2}\,\bar{f}_{0}\,v_{0}^{2}\,{\phi}_{q}} 
\end{align}

This can in principle be integrated numerically to arbitrary precision. But recalling that we have already parameterized the spectrum as a piecewise power-law, it is useful to parameterize other quantities such as $v$ and $\phi_{q}$ as approximate power-law functions of $p$ over the domain of the bin, e.g.\ take:
\begin{align}
\label{eqn:alpha.def} \alpha_{X} &\equiv \frac{\partial \ln{X}}{\partial \ln{p}}{\Bigr|}_{p=p_{0}} \approx \frac{\Delta \ln X}{\Delta \ln p} =  \frac{\ln{(X^{+}/X^{-})}}{\ln{(p^{+}/p^{-})}} 
\end{align} 
where $X^{\pm} \equiv X[p=p^{\pm}]$. So e.g.\ $\alpha_{q}=0$ exactly for $q=n$. For CRs with $p \gg m_{s}\,c$, $\alpha_{v} \approx 0$ and $\alpha_{q}\approx 1$ for $q=\epsilon$ (for $p \ll m_{s}\,c$, $\alpha_{v} \approx 1$ and $\alpha_{q}\approx2$ for $q=\epsilon$), so these are close to exact power-laws regardless, and so long as the spectral bins are small enough that there is no substantial spectral curvature within a bin (a necessary assumption for a piecewise-power-law treatment to be valid in the first place), approximating non-power-law behavior with Eq.~\ref{eqn:alpha.def} introduces no significant errors beyond our original piecewise power-law approximation.\footnote{More specifically, 
for the various $\alpha$ terms that appear in this paper, $\bar{f}_{0}$ and $\langle \mu\rangle$ or $\bar{f}_{1}$ within a bin are assumed to be exact power-laws by construction, so $\alpha_{f_{0}}$ and $\alpha_{\mu}$ have exact values but these can (and will) vary across cells and in time. The scattering rate $\nu$ is often assumed to be an exact power-law constant in time, but does not have to be (it could have curvature and/or vary with local plasma properties). Of course $\alpha_{q}=0$ identically for $q=n$, but for $q=\epsilon$, $\alpha_{q}$ is approximate (but it is fixed across all time and cells for a given bin. Likewise for $\alpha_{v}$). One could numerically evaluate all integrals presented here for the relevant $\omega$ terms exactly, without approximating terms such as $\epsilon$ as piecewise power-laws; but in our numerical tests this provides no appreciable improvement in accuracy compared to using the simpler, analytic power-law approximations we provide.} We can then immediately write: 
\begin{align}
\label{eqn:omega.delta.powerlaw} \omega_{\nabla,q} & \approx \frac{(3+\alpha_{f_{0}}+\alpha_{q})}{(3+\alpha_{f_{0}}+\alpha_{q}+2\,\alpha_{v})}\,\frac{(\xi^{3+\alpha_{f_{0}}+\alpha_{q}+2\,\alpha_{v}}-1)}{(\xi^{3+\alpha_{f_{0}}+\alpha_{q}}-1)}\,\xi^{-\alpha_{v}} \\ 
\nonumber &\approx 1 + \frac{\alpha_{v}}{6}\,\left( 3 + \alpha_{f_{0}} + \alpha_{q} + \alpha_{v} \right)\,\left| \ln{\xi} \right|^{2} + \mathcal{O}(|\ln{\xi}|^{4}) 
\end{align}
with $\xi \equiv p^{+}/p^{-} = \exp{(\ln{[p^{+}/p^{-}]})}$ (and the second expression above is a series expansion in $|\ln{\xi}|$). Note that with the definition in Eq.~\ref{eqn:alpha.def}, $\xi^{\alpha_{X}} \equiv X^{+}/X^{-}$ for any $X$, so we could equivalently write:
\begin{align}
\xi^{3+\alpha_{f_{0}}+\alpha_{q}+2\,\alpha_{v}} = \left(\frac{p^{+}}{p^{-}} \right)^{3}\,\left(\frac{\bar{f}_{0}^{+}}{\bar{f}_{0}^{-}} \right)\,\left(\frac{\psi_{q}^{+}}{\psi_{q}^{-}} \right)\,\left(\frac{v^{+}}{v^{-}} \right)^{2}
\end{align}
if the latter is more convenient.

With these assumptions, we can note\footnote{In this expression we also take $\bar{v}_{A}$ outside the integral. This depends implicitly on $p$ through $\nu_{\pm}$ as $\bar{v}_{A}=v_{A}\,(\bar{\nu}_{+}-\bar{\nu}_{-})/(\bar{\nu}_{+}-\bar{\nu}_{-})$. But like with $\chi$, this is almost always in one of two limits, either of which is $p$-independent. As shown in \citet{hopkins:2021.sc.et.models.incompatible.obs}, if extrinsic turbulence strongly dominates CR scattering and is forward/backward symmetric in the \Alf\ frame, then $\bar{v}_{A} \rightarrow 0$ is small and constant (and the term will be unimportant regardless). If otherwise (if e.g.\ self-confinement dominates or the scattering is asymmetric) then $\bar{v}_{A} = \pm v_{A}$ independent of $p$. So we can generally safely neglect the $p$-dependence of this within the bin here, especially as we later show that the steady-state behavior of this ``streaming'' term reduces to the bin-centered approximation.} $\mathcal{I}_{0,q} \approx \chi_{0}\,\alpha_{f_{0}}\,\bar{v}_{A,\,0}\, \int_{p^{-}}^{p^{+}}\,4\pi\,p^{2}\,dp\,\bar{\nu}\,\phi_{q}\,\bar{f}_{0}$, and immediately follow a similar procedure to obtain $\omega_{0}$: 
\begin{align}
\omega_{0,q} &\approx \frac{\int_{{p}^{-}}^{{p}^{+}}\,d{p}\,{p}^{2}\,\bar{\nu}\,\bar{f}_{0}\,{\phi}_{q}}{\int_{{p}^{-}}^{{p}^{+}}\,d{p}\,{p}^{2}\,\bar{\nu}_{0}\,\bar{f}_{0}\,\tilde{\phi}_{q}} \\
\nonumber & \approx \frac{(3+\alpha_{f_{0}}+\alpha_{q})}{(3+\alpha_{f_{0}}+\alpha_{q}+\alpha_{\nu})}\,\frac{(\xi^{3+\alpha_{f_{0}}+\alpha_{q}+\alpha_{\nu}}-1)}{(\xi^{3+\alpha_{f_{0}}+\alpha_{q}}-1)}\,\xi^{-\alpha_{\nu}/2} \\ 
\nonumber &\approx 1 + \frac{\alpha_{\nu}}{12}\,\left( 3 + \alpha_{f_{0}} + \alpha_{q} +\alpha_{\nu}/2 \right)\,\left| \ln{\xi} \right|^{2} + \mathcal{O}(|\ln{\xi}|^{4}) 
\end{align}

Note that $\alpha_{\nu}$ corresponds to $\bar{\nu} \propto p^{\alpha_{\nu}}$; for the commonly-adopted phenomenological assumption in modeling Galactic and Solar System CR observables that the diffusivity scales as $\kappa \propto {R}^{\delta}$ for CR rigidity ${R}$ (at energies where $\beta\approx1$), we have $\alpha_{\nu} \approx -\delta$, so those observations imply $-0.7 \lesssim \alpha_{\nu} \lesssim -0.4$ \citep{blasi:cr.propagation.constraints,vladimirov:cr.highegy.diff,gaggero:2015.cr.diffusion.coefficient,2016ApJ...819...54G,2016ApJ...824...16J,cummings:2016.voyager.1.cr.spectra,2016PhRvD..94l3019K,evoli:dragon2.cr.prop,2018AdSpR..62.2731A,delaTorre:2021.dragon2.methods.new.model.comparison,hopkins:cr.multibin.mw.comparison}.

\subsubsection{Terms which Depend Strongly on Pitch Angle}

Now consider $\omega_{1,q}$. Here, we cannot neglect the implicit $\mu$-dependence, because the fluxes $F_{q}$ are directly proportional to $\langle \mu \rangle$. So make the {\em ansatz}, like above, that we can approximate $\langle \mu \rangle \propto p^{\alpha_{\mu}}$ over the (relatively narrow) width of the bin, giving: 
\begin{align}
\label{eqn:omega.1} \omega_{1,q} &= \frac{\int_{{p}^{-}}^{{p}^{+}}\,d{p}\,{p}^{2}\,\bar{\nu}\,{\langle \mu \rangle}\,{v}\,\bar{f}_{0}\,{\phi}_{q}}{\int_{{p}^{-}}^{{p}^{+}}\,d{p}\,{p}^{2}\,\bar{\nu}_{0}\,{\langle \mu \rangle}\,{v}\,\bar{f}_{0}\,{\phi}_{q}}  \\ 
\nonumber & \approx \frac{(3+\alpha_{f_{0}}+\alpha_{q}+\alpha_{v}+\alpha_{\mu})}{(3+\alpha_{f_{0}}+\alpha_{q}+\alpha_{v}+\alpha_{\mu}+\alpha_{\nu})}\, \\ 
\nonumber & \ \ \ \ \ \times \frac{(\xi^{3+\alpha_{f_{0}}+\alpha_{q}+\alpha_{v}+\alpha_{\mu}+\alpha_{\nu}}-1)}{(\xi^{3+\alpha_{f_{0}}+\alpha_{q}+\alpha_{v}+\alpha_{\mu}}-1)}\,\xi^{-\alpha_{\nu}/2} \\ 
\nonumber &\approx 1 + \frac{\alpha_{\nu}}{12}\,\left( 3 + \alpha_{f_{0}} + \alpha_{q}  +\alpha_{v}+\alpha_{\mu} +\alpha_{\nu}/2 \right)\,\left| \ln{\xi} \right|^{2} \\ 
\nonumber & \ \ \ \ \ + \mathcal{O}(|\ln{\xi}|^{4}) 
\end{align}
Here, as in the expressions above and in various expressions below, the first (explicit integral) expression is exact. The second makes the power-law substitution, and is exact to the extent that the power-law approximation for the quantities inside the integrand is exact over the width of the bin.\footnote{Technically we have to be careful about the case where the integrand with $dp$ scales exactly as $p^{-1}$, in which case the power-law expressions should evaluate to $\ln$ instead of those shown. But for any case where the index is not exactly negative one this is can be solved without issue and if constructing a numerical interpolation one can interpolate across this boundary without divergences.} The third is a series approximation in $|\ln{\xi}|$, which is generally not necessary for our numerical evaluations in a code implementation of these methods, but is convenient here for intuition-building and understanding different limits discussed below.

Eq.~\ref{eqn:omega.1} would allow us to evolve Eq.~\ref{eqn:DtF.familiar}, except now we have introduced a new parameter $\alpha_{\mu}$ which is not {\em a priori} specified. However, it is not actually the case that $\alpha_{\mu}$ is unconstrained. Since our update to the DF (Eq.~\ref{eqn:f0.power.law}) requires evolving both of a pair $(q,\,q^{\prime})$ ($=(n,\,\epsilon)$) with associated fluxes ($F_{q}$, $F_{q^{\prime}}$), then by combining the definitions of $(q,\,q^{\prime},\,F_{q},\,F_{q^{\prime}})$, one can show there is one independent consistency relation that must be satisfied:
\begin{align}
\label{eqn:Psi} \Psi_{qq^{\prime}} &\equiv {\left(\frac{F_{q}}{q\,v_{0}} \right)}{\Big/}{\left(\frac{F_{q^{\prime}}}{q^{\prime}\,v_{0}} \right)} = \frac{q^{\prime}\,F_{q}}{q\,F_{q^{\prime}}} \\ 
\nonumber &\equiv \frac{\left( \int_{{p}^{-}}^{{p}^{+}}\,d{p}\,{p}^{2}\,{\langle \mu \rangle}\,{v}\,\bar{f}_{0}\,{\phi}_{q} \right)}{\left( \int_{{p}^{-}}^{{p}^{+}}\,d{p}\,{p}^{2}\,{\langle \mu \rangle}\,{v}\,\bar{f}_{0}\,{\phi}_{q}^{\prime} \right)} 
\frac{\left( \int_{{p}^{-}}^{{p}^{+}}\,d{p}\,{p}^{2}\,\bar{f}_{0}\,{\phi}_{q^{\prime}} \right)}{\left( \int_{{p}^{-}}^{{p}^{+}}\,d{p}\,{p}^{2}\,\bar{f}_{0}\,{\phi}_{q} \right)} \\ 
\nonumber &\approx 
\frac{(3+\alpha_{f_{0}}+\alpha_{q})}{(3+\alpha_{f_{0}}+\alpha_{q^{\prime}})}\,
\frac{(3+\alpha_{f_{0}}+\alpha_{q^{\prime}}+\alpha_{v}+\alpha_{\mu})}{(3+\alpha_{f_{0}}+\alpha_{q}+\alpha_{v}+\alpha_{\mu})}\, \\ 
\nonumber & \ \ \ \ \ \times \ 
\frac{(\xi^{3+\alpha_{f_{0}}+\alpha_{q^{\prime}}}-1)}{(\xi^{3+\alpha_{f_{0}}+\alpha_{q}}-1)}
\frac{(\xi^{3+\alpha_{f_{0}}+\alpha_{q}+\alpha_{v}+\alpha_{\mu}}-1)}{(\xi^{3+\alpha_{f_{0}}+\alpha_{q^{\prime}}+\alpha_{v}+\alpha_{\mu}}-1)}  \\
\nonumber &\approx 1 + \frac{1}{12}\,\left(\alpha_{q}-\alpha_{q^{\prime}}\right)\,\left(\alpha_{v} + \alpha_{\mu}\right)\,\left| \ln{\xi} \right|^{2} + \mathcal{O}(|\ln{\xi}|^{4})
\end{align}
Once again we give the exact integrals, solution making the power-law replacement, and series approximation in turn.

This is sufficient to specify $\alpha_{\mu}$ and therefore $\omega_{1,q}$, according to the different integration methods described below.

\subsubsection{Solution Methods}
\label{sec:solution.methods}

With expressions for $\omega$, Eq.~\ref{eqn:DtF.familiar} can be numerically integrated with exactly the same numerical methods as used for the ``bin-centered'' method above -- the $\omega$ terms only amount to a scalar renormalization of $\mathbb{D}_{0}$, $\bar{\nu}_{0}$, and $v_{\rm st}$ which are arbitrary anyways from the point of view of the numerical method. The added complication comes almost entirely from determining $\alpha_{\mu}$ consistently to evaluate these terms. Consider three methods to do so:

\begin{enumerate}
\item{Exact:} One option is to exactly update $(q,\,q^{\prime},\,F_{q},\,F_{q^{\prime}})$ subject to the constraint $\Psi_{qq^{\prime}}$ (Eq.~\ref{eqn:Psi}). One can think of this as ``replacing'' the value of $\alpha_{\mu}$ with that determined by $\Psi_{qq^{\prime}}$ in the original equations for $(q,\,q^{\prime},\,F_{q},\,F_{q^{\prime}})$. While do-able in principle, this (a) is extremely non-linear and involves inverting several complicated and numerically stiff functions of four variables; (b) couples the $(q,\,q^{\prime},\,F_{q},\,F_{q^{\prime}})$ variables explicitly so we are forced to update all simultaneously with a single implicit step, i.e.\ we cannot operator-split as is usually desired; and (c) can sometimes lead to non-invertible expressions if great care is not taken with numerical errors.

\item{Approximate, Integrated:} Alternatively, if the numerical method explicitly integrates the variables $F_{q}$ and $F_{q^{\prime}}$ (e.g.\ two-moment methods), then we can insert the values of $(q,\,q^{\prime},\,F_{q},\,F_{q^{\prime}})$ at some point in the timestep (at the beginning of the step or ``drifted'' to a half-step for a standard explicit method, or their exact values at step-end for implicit integration) into Eq.~\ref{eqn:Psi} and solve for $\alpha_{\mu}$ from that expression, then use this value of  $\alpha_{\mu}$ in Eq.~\ref{eqn:DtF.familiar} to calculate the update to $F_{q}$ and $F_{q^{\prime}}$. This is similar to how the other variables in Eq.~\ref{eqn:DtF.familiar} appear and is numerically straightforward (the single numerical inversion of Eq.~\ref{eqn:Psi} for a given $\Psi_{qq^{\prime}}$ value is straightforward as well). We find this works quite well.\footnote{Some numerical caution is still always needed. For example, if one adopts the power-law approximations given above, then one needs to treat the regime around certain values where some expressions would seemingly produce divergences carefully. Specifically, this arises when the integrand in the original exact expression takes values $\sim \int p^{-1}\,dp$, so the power-law solutions should be replaced with logarithmic solutions: for example if $3+\alpha_{f_{0}}+\alpha_{q} = 0$ in Eq.~\ref{eqn:Psi}, which could numerically give a $0/0$ error. For power-law indices close to these critical values we recommend either using the exact integral solutions (ideally), or a loopkup table designed to be interpolated over the relevant range, rather than taking the power-law expressions directly at face value.} And in one-moment methods, we must determine the appropriate $F_{q}$, $F_{q^{\prime}}$ self-consistently and simultaneously, which we discuss below.
\item{``Local-Steady-State'' Values:} A still simpler, but even more numerically robust method is to not solve for $\alpha_{\mu}$ from the constraint Eq.~\ref{eqn:Psi} exactly, but to instead adopt the value it {\em would} have for the corresponding terms in Eq.~\ref{eqn:DtF.familiar} if the flux equation were in local steady-state. We derive this and further define below. This has the advantage that it is extremely robust and trivially numerically stable (provided whatever integration method used for the ``bin centered'' approximation is also stable). It sacrifices manifest consistency between Eq.~\ref{eqn:Psi} and Eq.~\ref{eqn:DtF.familiar} for $q,\,q^{\prime}$, but we are guaranteed that when the flux equations $D_{t}F_{q}$ are close to local steady-state (which is usually the case), the consistency relations are satisfied.

\end{enumerate}

We also note that while it is generally advisable to use the full numerical expressions for $\omega$, the series expressions we show (expansions in $|\ln{\xi}|$) work surprisingly well for even large $\xi$, valid to better than $\sim 10\%$ for all $\omega$ terms for any $\xi \lesssim 3$, and for some of the terms (especially in the ultra-relativistic limit) the series expression works well up to $\xi \lesssim 100$ (assuming the underlying terms could, in fact, be approximated as power-laws reliably over that dynamic range).

\subsection{Local Flux Steady-State Behaviors}
\label{sec:local.flux.steady}

Consider the case where the flux equations (Eq.~\ref{eqn:spatial.flux}) reach approximate local-steady-state, i.e.\ $|D_{t} F_{q}| \rightarrow 0$ (or $|D_{t} F_{q}| \ll |\nu\,F_{q}|$). This occurs on approximately the scattering time $\sim \nu^{-1}$, which is very short in the Galactic ISM (from observations, $\nu^{-1} \sim 30\,{\rm yr}$ for $\sim 1\,$GeV CRs; see \citealt{hopkins:cr.multibin.mw.comparison}). Thus even if we explicitly evolve $F_{q}$, we expect it to be close to this ``local flux steady-state'' value in many regimes. Moreover, the ``one-moment'' numerical methods assume this is {\em exactly} true, to directly solve for $F_{q}$ and insert it into Eq.~\ref{eqn:spatial} to  directly obtain a diffusion-streaming equation for the CRs \citep[see e.g.][and references therein]{Zwei13}. 
Noting that this implies the strong-scattering limit, so the CRs are nearly-isotropic ($\chi\rightarrow 1/3$, $\mathbb{D} \rightarrow \mathbb{I}/3$), we immediately obtain from Eq.~\ref{eqn:DtF.familiar}: 
\begin{align}
\label{eqn:F.steady} F_{q} &\rightarrow {v}^{\rm eff}_{{\rm st},q}\,q - \frac{v_{0}^{2}}{3\,{\nu}^{\rm eff}_{0,q}}\,\nabla_{\|} \left( \omega_{\nabla,q}\,q \right)
\end{align}
where $\nabla_{\|} \equiv \bhat \cdot \nabla$. 
So up to the ``effective'' coefficients being slightly modified by the $\omega$ terms, this is just the usual steaming/diffusion expression, with streaming speed ${v}^{\rm eff}_{{\rm st},q}$ and effective {\em anisotropic} diffusivity $\kappa_{\|} \sim {v_{0}^{2}}/{3\,{\nu}^{\rm eff}_{0,q}}$ (if we assume isotropically tangled magnetic fields on small scales, this can be further approximated as an isotropic diffusivity $D_{0} \sim \kappa_{\|}/3$).

\subsubsection{The ``\Alf{ic} Streaming-Dominated'' Limit}
\label{sec:local.flux.steady.streaming}

Consider the case where the \Alf{ic} streaming term dominates in Eq.~\ref{eqn:F.steady}, $F_{q} \rightarrow {v}^{\rm eff}_{{\rm st},q}\,q$ (this can occur in e.g.\ self-confinement models when $\bar{\nu} \rightarrow \infty$). Then the Eq.~\ref{eqn:Psi} becomes $\Psi_{qq^{\prime}} = (\omega_{0,q}\,\omega_{1,q^{\prime}})/(\omega_{0,q^{\prime}}\,\omega_{1,q})$. This is solved exactly if and only if $\alpha_{\mu} \rightarrow -\alpha_{v}$, i.e.\ the CR drift velocity $v_{\rm drift} = \langle \mu\rangle\,v \propto p^{0}$ is independent of momentum (as it must be, since they are drifting, by definition in this limit, at the momentum-independent streaming speed across the bin). Inserting this into the expressions for $v_{{\rm st},q}^{\rm eff}$, we immediately have:
\begin{align}
\frac{\omega_{0,q}}{\omega_{1,q}} & \rightarrow 1,\ {v}^{\rm eff}_{{\rm st},q} \rightarrow -\frac{\alpha_{f_{0}}}{3}\,\bar{v}_{A,0}
\end{align}
In this limit, because the drift velocity is constant (across the bin), and the gradient/$\mathbb{D}$ term and diffusive terms are irrelevant, we see that we have recovered {\em exactly} the same $F_{q}$ that we would have in the bin-centered approximation.

\subsubsection{The Diffusive or Super-\Alf{ic} Limit}
\label{sec:local.steady.state.diffusion}

Now consider the limit where the ``diffusive'' term dominates in Eq.~\ref{eqn:F.steady}, so $F_{q} \rightarrow (v_{0}^{2}/3\,{\nu}^{\rm eff}_{0,q})\,\nabla_{\|} ( \omega_{\nabla,q}\,q )$. Note that when some literature refers to ``super-\Alf{ic} streaming,'' this still comes from this particular term (and there is no distinction, for our purposes here). The constraint equation then becomes $\Psi_{qq^{\prime}} = (\omega_{1,q^{\prime}}\,\ell_{q^{\prime}})/(\omega_{1,q}\,\ell_{q})$, where
\begin{align}
\ell_{q} &\equiv \frac{q}{\nabla_{\|}(\omega_{\nabla,q}\,q)}
\end{align}
Solving for $\alpha_{\mu}$ from this constraint gives a highly nonlinear equation to be solved for $\alpha_{\mu} \rightarrow \alpha_{\mu}(\ell_{q^{\prime}}/\ell_{q},\,\alpha_{\nu},\,\alpha_{q},\,\alpha_{v},\,\alpha_{\mu}\,\alpha_{q^{\prime}})$,\footnote{If we take $\alpha_{\mu} \rightarrow -\alpha_{v} - \alpha_{\nu} + \Delta \alpha_{\mu}$ and, for compactness, write $\alpha_{3\alpha_{f_{0}} q} \equiv 3+\alpha_{f_{0}}+\alpha_{q}$, we have: 
\begin{align}
\label{eqn:Lq.footnote.1} \frac{\ell_{q^{\prime}}}{\ell_{q}} &\approx \frac{\alpha_{3\alpha_{f_{0}} q}\,(\alpha_{3\alpha_{f_{0}} q^{\prime}}+ \Delta \alpha_{\mu})}{\alpha_{3\alpha_{f_{0}} q^{\prime}}\,(\alpha_{3\alpha_{f_{0}} q}+ \Delta \alpha_{\mu})}\,\frac{(\xi^{\alpha_{3\alpha_{f_{0}} q^{\prime}}}-1)\,(\xi^{\alpha_{3\alpha_{f_{0}} q}+\Delta \alpha_{\mu}}-1)}{(\xi^{\alpha_{3\alpha_{f_{0}} q}}-1)\,(\xi^{\alpha_{3\alpha_{f_{0}} q^{\prime}}+\Delta \alpha_{\mu}}-1)}
\end{align}
which can then be solved for $\Delta \alpha_{\mu}$.}
It is more instructive to parameterize $\ell_{q}$ in a similar piecewise-power-law manner: let us define $\nabla_{\|}\,\bar{f}_{0} \equiv \bar{f}_{0}/\ell_{f}$ where $\ell_{f}=\ell_{f}(p)$ is defined over an infinitesimally small range of $p$, and let us assume this scales similarly as $\ell_{f} \propto p^{\alpha_{\ell}}$.\footnote{We can also immediately calculate the relation between $\alpha_{\ell}$  and $\ell_{q^{\prime}}/\ell_{q}$: 
\begin{align}
\nonumber \frac{\ell_{q^{\prime}}}{\ell_{q}} &\approx \frac{\alpha_{3\alpha_{f_{0}} q}\,(\alpha_{3\alpha_{f_{0}} q^{\prime}}+ 2\,\alpha_{v}-\alpha_{\ell})}{\alpha_{3\alpha_{f_{0}} q^{\prime}}\,(\alpha_{3\alpha_{f_{0}} q}+ 2\,\alpha_{v}-\alpha_{\ell})}\,\frac{(\xi^{\alpha_{3\alpha_{f_{0}} q^{\prime}}}-1)\,(\xi^{\alpha_{3\alpha_{f_{0}} q}+2\,\alpha_{v}-\alpha_{\ell}}-1)}{(\xi^{\alpha_{3\alpha_{f_{0}} q}}-1)\,(\xi^{\alpha_{3\alpha_{f_{0}} q^{\prime}}+2\,\alpha_{v}-\alpha_{\ell}}-1)} \\ 
\label{eqn:Lq.footnote.2} &\approx 1 + \frac{1}{12}\,\left(\alpha_{q}-\alpha_{q^{\prime}}\right)\,\left(2\,\alpha_{v} - \alpha_{\ell}\right)\,\left| \ln{\xi} \right|^{2} + \mathcal{O}(|\ln{\xi}|^{4})
\end{align}
which allows us to solve for $\alpha_{\ell}$. 
}
If we combine this with the steady-state expressions for $F_{q}$ in terms of the relevant gradients, and use Eqs.~\ref{eqn:Lq.footnote.1}-\ref{eqn:Lq.footnote.2}, we see that the consistency relations are satisfied exactly for $\alpha_{\mu} \rightarrow -\alpha_{\nu} + \alpha_{v} - \alpha_{\ell}$, which we can immediately insert in Eq.~\ref{eqn:omega.1}.

With these definitions and some similar algebra, it is also convenient to note that we can write: 
\begin{align}
\label{eqn:Fq.kappa} F_{q} &\rightarrow -\kappa_{\|,q}^{\ast}\,\nabla_{\|} q = -\left( \omega_{\kappa,q}\,\frac{v_{0}^{2}}{3\,\bar{\nu}_{0}} \right)\,\nabla_{\|} q \\ 
\label{eqn:omega.kappa} \omega_{\kappa,q} &\approx
\frac{(3 + \alpha_{f_{0}} + \alpha_{q} - \alpha_{\ell})}{(3 + \alpha_{f_{0}} + \alpha_{q} - \alpha_{\ell} + 2\,\alpha_{v} - \alpha_{\nu})} \, \\
\nonumber &\ \ \ \ \ \times\ \frac{(\xi^{3 + \alpha_{f_{0}} + \alpha_{q} - \alpha_{\ell} + 2\,\alpha_{v} - \alpha_{\nu}}-1)}{(\xi^{3 + \alpha_{f_{0}} + \alpha_{q} - \alpha_{\ell}}-1)}\,\xi^{-\alpha_{v} + \alpha_{\nu}/2} \\ 
\nonumber &\approx 1 + \frac{(2\,\alpha_{v} - \alpha_{\nu} )}{12}\,\left(3+\alpha_{f_{0}}+\alpha_{q} + \alpha_{v} - \alpha_{\ell} - \alpha_{\nu}/2 \right)\,\left| \ln{\xi} \right|^{2} \\ 
\nonumber & \ \ \ \ \ + \mathcal{O}(|\ln{\xi}|^{4}) 
\end{align}
i.e.\ the effective diffusivity is simply modified by a correction factor $\omega_{\kappa,q}$. For $\gtrsim$\,GeV CRs, where we have empirically typical $\alpha_{f_{0}} \sim -4.7$ (from direct observation; e.g.\ \citealt{cummings:2016.voyager.1.cr.spectra}), $\alpha_{\nu} \sim -0.6$ (from modeling of primary-to-secondary ratios and similar constraints; \citealt{delaTorre:2021.dragon2.methods.new.model.comparison,hopkins:cr.multibin.mw.comparison,korsmeier:2021.light.element.requires.halo.but.upper.limit.unconfined}), $\alpha_{v} \sim 0$ (from the fact that these are ultra-relativistic), $\alpha_{\ell} \lesssim 0.1$ (from modeling spatially-resolved Galactic $\gamma$-ray profiles at different energies; e.g.\ \citealt{tibaldo.2015:diffuse.gamma.ray.cr.profile.constraints,acero:2016.gamma.ray.constraints.cr.emissivity,yang.2016:diffuse.gamma.ray.cr.profile.constraints,hopkins:cr.multibin.mw.comparison}), we obtain $\omega_{\kappa,q}-1 \sim 0.05\,(1.4-\alpha_{q}+\alpha_{\ell})\,|\ln{\xi}|^{2} + \mathcal{O}(\,|\ln{\xi}|^{4}) \sim -(0.08,\,0.03)\,|\ln{\xi}|^{2}$ for $q=(n,\,\epsilon)$. The ``mean'' correction (both are $<0$ because for these energies, most of the CR number and  energy is biased towards the lower-$p$ end of the ``bin,'' where the effective $\kappa$ is smaller) is modest and not so important, given the (very large) systematic theoretical uncertainties in the ``correct'' scaling of $\bar{\nu}_{0}$ \citep{Zwei13,zweibel:cr.feedback.review,farber:decoupled.crs.in.neutral.gas,yan.lazarian.04:cr.scattering.fast.modes,yan.lazarian.2008:cr.propagation.with.streaming,holguin:2019.cr.streaming.turb.damping.cr.galactic.winds,bustard:2020.crs.multiphase.ism.accel.confinement,hopkins:2020.cr.transport.model.fx.galform,hopkins:cr.transport.constraints.from.galaxies,hopkins:cr.multibin.mw.comparison}. What is important is the relative correction: the CR number flux is more strongly modified (because CR number is more strongly dominated by the low-$p$ end of the bin), and the (small) difference here causes the spectral slope to steepen {\em within the bin} as CRs diffuse.

Note that if we must still evaluate $\ell_{q^{\prime}}/\ell_{q}$ to determine $\alpha_{\ell}$ for Eq.~\ref{eqn:Fq.kappa} above, then it is not necessarily more computationally useful than just using $F_{q} \rightarrow (v_{0}^{2}/3\,{\nu}^{\rm eff}_{0,q})\,\nabla_{\|} ( \omega_{\nabla,q}\,q )$ as we would have previously, but it is still useful to guide our intuition. Moreover, we can note that in the limit where the diffusive term dominates the flux, with negligible losses, and the CR $(n,\,\epsilon)$ equations are themselves close to steady-state (assuming also $\bar{\nu}$ and the source injection spectrum do not vary strongly with spatial location), then $\alpha_{\ell} \rightarrow 0$. Since that is precisely the regime where it matters most to get this correction ``right,'' we can assume this without much loss of accuracy given our other significant simplifications above.

\subsubsection{The ``Local-Steady-State'' Approximation for Flux Corrections}
\label{sec:local.steady.state}

With all this in mind, if one adopts a  two-moment method (evolving $F_{q}$ explicitly) with the primary goal of capturing the exact behavior in the three possible limits of Eq.~\ref{eqn:spatial.flux} (free-streaming/weak-scattering, or near-isotropic/strong-scattering/diffusive, or trapped/advective/\Alf{ic}-streaming [$\bar{\nu}\rightarrow \infty$]),\footnote{Even relatively sophisticated closure schemes for evolving $\langle \mu^{2} \rangle$ proposed in the literature focus primarily on the behavior in these three limits, as opposed to intermediate cases; see \citet{hopkins:m1.cr.closure} for a review.} then it is sufficient to adopt the ``local-steady-state'' approximation for $\alpha_{\mu}$ in Eq.~\ref{eqn:DtF.familiar} using the appropriate value of $\alpha_{\mu}$ each term would have if it were dominant. This gives:
\begin{align}
D_{t} F_{q} &+ v_{0}^{2}\,\bhat \cdot \nabla \cdot \left( \mathbb{D}_{0}\,\omega_{\nabla,q}\,q \right) = -\nu^{\ast}_{q}\,\left( F_{q} - v_{{\rm st},0}\,q \right) \\ 
\nonumber v_{{\rm st},0} &\equiv -\chi_{0}\,\alpha_{f_{0}}\,\bar{v}_{A,0} \\
\nonumber \nu^{\ast}_{q} &\equiv \bar{\nu}_{0}\,\omega_{1,q}[ \alpha_{\mu} \rightarrow \alpha_{v} - \alpha_{\nu} - \alpha_{\ell} ]  
\end{align}
(with $\omega_{\nabla,q}$ from Eq.~\ref{eqn:omega.delta.powerlaw} and $\omega_{1,q}$ from Eq.~\ref{eqn:omega.1}). 
One can immediately verify this reduces correctly to any of the relevant local-steady-state limits above. 

If one evolves a ``one-moment'' method -- e.g.\ evolving the CRs according to a single streaming+diffusion or Fokker-Planck type approximation (valid only in the strong-scattering limits), then we can approximate the limits of interest via:
\begin{align}
F_{q} \rightarrow -\frac{\alpha_{f_{0}}}{3}\,\bar{v}_{A,0}\,q - \left( \omega_{\kappa,q}\,\frac{v_{0}^{2}}{3\,\bar{\nu}_{0}} \right)\,\nabla_{\|} q
\end{align}
(with $\omega_{\kappa,q}$ from Eq.~\ref{eqn:omega.kappa}) 
where $\alpha_{\ell}$ in $\omega_{\kappa,q}$ can be computed or (for even greater simplicity), approximated as $\approx0$ without severe loss of accuracy.

\begin{figure}
	\includegraphics[width=0.98\columnwidth]{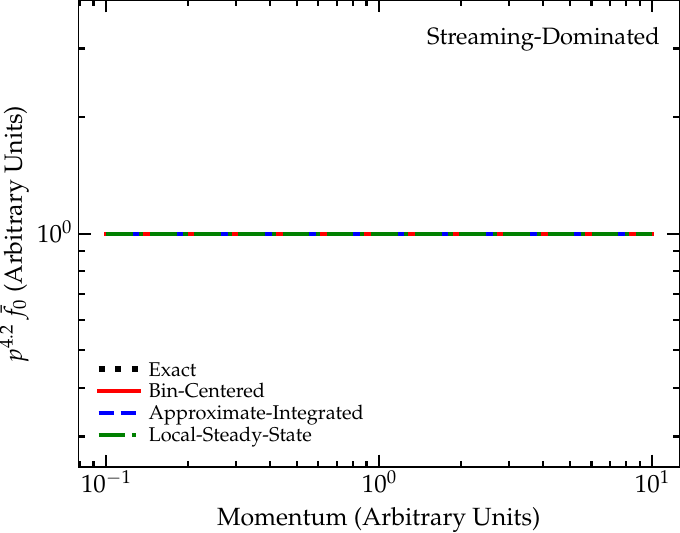} \\ 
	\includegraphics[width=0.98\columnwidth]{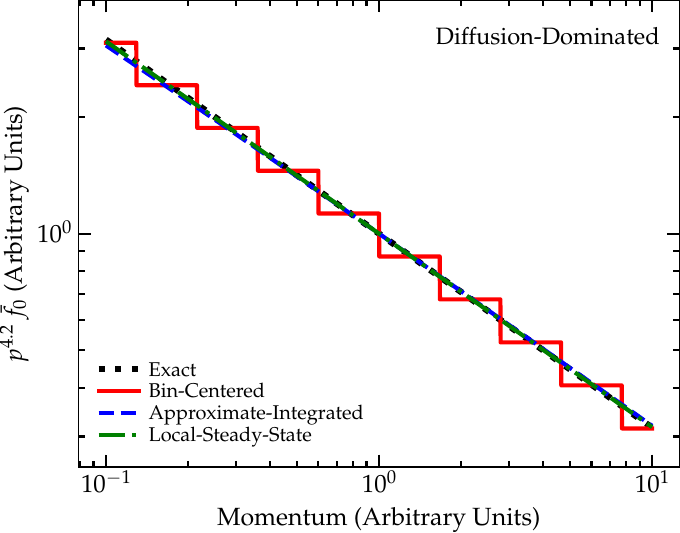} \\ 
	\vspace{-0.2cm}
	\caption{Numerical tests (\S~\ref{sec:test}) of our proposed correction terms for spatial transport of piecewise-power-law spectra. We consider a homogeneous, one-dimensional stratified atmosphere with a continuous injection spectrum $j_{\rm inj} \propto p^{-4.2}$ at the lower boundary and outflow from the upper boundary, constant streaming speed $\bar{v}_{A}$ and scattering rate $\bar{\nu} \propto p^{-0.5}$, discretized into $10$ momentum intervals, and evolved until steady-state using the numerical methods described in the text.
	We compare exact analytic steady-state solutions to numerical solutions using either (1) the ``bin-centered'' approximation ($\omega_{\nabla}=\omega_{0}=\omega_{1}=1$; \S~\ref{sec:bin.centered}), (2) the ``approximate-integrated'' method (\S~\ref{sec:solution.methods}) to solve for the $\omega$ terms (given the in-code evolved values of $n,\,\epsilon,\,F_{n},\,F_{\epsilon}$ to calculate $\alpha_{\nu}$ from Eq.~\ref{eqn:Psi}), and (3) the ``local-steady-state'' approximation for $\omega$ terms (\S~\ref{sec:local.steady.state}), using the local-steady-state values of $\alpha_{\mu}$. 	
	{\em Top:} Parameters chosen so the transport is streaming-dominated ($\bar{\nu}$ very large). We plot the steady-state spectrum compensated by $p^{4.2}$ and in units such that the exact solution equals unity. In the streaming-dominated limit, the transport speed is momentum-independent so the spectrum is simply advected without change in spectral slope, and the different approximations behave identically.
	{\em Bottom:} Parameters chosen so the transport is diffusion-dominated ($\bar{v}_{A}=0$). The ``bin-centered'' approximation introduces well-known step artifacts, as a result of assuming the scattering rate and $\langle \mu \rangle$ are constant within each bin, which conserves the injection slope within each momentum bin. Both our proposed methods for including the $\omega$ terms produce the correct spectral slopes within bins.
	\vspace{-0.2cm}
	\label{fig:numerical.test}}
\end{figure}

\section{Simple Numerical Tests}
\label{sec:test}

In Fig.~\ref{fig:numerical.test}, we consider a simple illustrative numerical test of the proposed methods. To isolate the interesting behavior and construct a simple, analytically-tractable test problem, we consider transport of a power-law injection spectrum in a plane-parallel atmosphere, analogous to classic thin disk or leaky-box type models for CRs. Specifically, consider an infinitely-thin source plane in the $xy$ axis, in a homogeneous, stationary background (e.g.\ ${\bf v}_{\rm gas}=\mathbf{0}$, $\bhat=\hat{z}=\,$constant) with space-and-time-independent $\bar{v}_{A}=$\,constant and $\bar{\nu} \propto p^{-0.5}$, $T \propto p$ and $\beta \approx 1$ (e.g.\ the ultra-relativistic limit, though this choice has no effect on our conclusions), ignoring all non-spatial transport terms (e.g.\ catastrophic or radiative losses) except for injection in the source plane at a constant rate per unit area $J_{0} \equiv dN_{\rm cr}/dt\,dA\,d^{3}{\bf p}$. Numerically, we integrate this on a domain with $10$ spatial cells in the vertical direction from $z=0$ (with an inflow/injection boundary) to $z=1$ (with an outflow boundary) in arbitrary code units, and injection slope $j_{\rm inj} \propto p^{-4.2}$ similar to physically-expected values, using the finite-volume two-moment method (evolving $n,\,\epsilon,\,F_{n},\,F_{\epsilon}$) in the code {\small GIZMO} \citep{hopkins:gizmo,hopkins:mhd.gizmo,hopkins:gizmo.diffusion,hopkins:cr.multibin.mw.comparison},\footnote{We have also tested these problems implementing the 10-element discretization in 1D, solved via a Crank-Nicholson scheme in Python using either the two-moment equations or (since we consider the steady-state solutions) directly integrating the single-moment streaming+diffusion equation in \S~\ref{sec:local.flux.steady}, which gives indistinguishable results to those shown in Fig.~\ref{fig:numerical.test}.} with the $\omega$ values determined according to the different proposed methods described in the text. We discretize the momentum domain with $10$ bins over $2\,$dex (though again, given the simplifications of our problem, the dynamic range of $p$ is not important to our conclusions). We set the normalization of $\bar{v}_{A}$ and $\bar{\nu}$ to two different values to compare two limits.

First, we consider a ``streaming-dominated'' limit, obtained by setting $\bar{\nu}$ to a very large value ($\sim 10^{6}\,p^{-0.5}$ in code units) with $\bar{v}_{A}=1$ (and effective diffusion coefficient $v^{2}/3\,\bar{\nu}$ set to an arbitrarily small value), so analytically $\bar{f}_{1} \rightarrow (\tilde{D}_{\mu\mu}/\tilde{D}_{\mu p})\,\partial_{p}\,\bar{f}_{0}$. This has a simple constant-flux steady-state solution with $v\,\bar{f}_{1}=-\bar{v}_{A}\,\alpha_{f_{0}}\,\bar{f}_{0} = J_{0}$, so $\bar{f}_{0} \rightarrow J_{0}/|\alpha_{f_{0}}\,\bar{v}_{A}|$ is spatially-uniform and proportional to the injection spectrum (i.e.\ $\alpha_{f_{0}} = \alpha_{j_{\rm inj}} = -4.2$). As predicted in \S~\ref{sec:local.flux.steady.streaming}, the injection spectrum is simply advected here, so all methods (including the simple bin-centered approximation) reproduce the exact solution in Fig.~\ref{fig:numerical.test} in this limit.

Second, we consider a ``diffusion-dominated'' case, setting $\bar{v}_{A}=0$ with finite $\bar{\nu}$ (evolved to several times the effective diffusion time). In steady-state now $v\,\bar{f}_{1} = -(v^{2}/3\,\bar{\nu})\,\partial_{z}\,\bar{f}_{0} =  J_{0}$, so $\partial_{z}\,\bar{f}_{0} = -(3\,\bar{\nu}/v^{2})\,J_{0}=\,$constant in space and $\propto p^{-0.5}$. Because higher-energy CRs have a lower $\bar{\nu}$, and correspondingly larger effective diffusivity $v^{2}/3\,\bar{\nu}$, they escape faster and their steady-state abundance (relative to injection) is reduced, steepening the spectrum by one power of $\bar{\nu}$. All the numerical methods in Fig.~\ref{fig:numerical.test} capture this effect ``on average'' across bins. But for the ``bin-centered'' method, as anticipated in \S~\ref{sec:bin.centered}, we effectively ignore the variation of $\bar{\nu}$ {\em within} each bin (taking the bin-centered $\bar{\nu}_{0}$ as constant across each bin). This means we very slightly over-estimate the total value of $\bar{\nu}$ (leading to a small under-estimate of the mean $\bar{f}_{0}$, averaged over the bin), but more importantly the method conserves the spectral slope within each bin, producing the ``step'' structures seen. On the other hand, introducing the scalar $\omega$ correction terms as proposed in this paper, with either method in Fig.~\ref{fig:numerical.test}, leads to excellent agreement with the exact solutions (with the slope in each bin numerically agreeing with the exact solution to better than $\sim 1\%$).

\section{Applications to Radiation/Neutrino Dynamics}

It is natural to ask whether the methodology above can be cross-applied to radiation or neutrino transport, where one can easily imagine situations in which a similar piecewise-power-law reconstruction of the radiation spectrum would be useful.

For the sake of consistency with the large radiation/neutrino transport literature, in this section we will consider a different set of variable definitions matching the convention in those fields. Let $\nu$ refer to the radiation frequency (so $h\nu$ is energy, analogous to $p$ for CRs), so the specific intensity $I_{\nu}({\bf n},\,\nu,\,{\bf x},\,t)$ is equivalent to the DF $f$ in terms of the radiation direction unit vector ${\bf n}$, the mean/isotropic intensity $J_{\nu}\equiv (4\pi)^{-1}\,\int I_{\nu}\,d\Omega$ is analogous to $\bar{f}_{0}$, $\mathbb{D}_{\nu}$ ($\equiv (4\pi\,J_{\nu})^{-1}\,\int d\Omega\,{\bf  n} \otimes {\bf n}\,I_{\nu} $) is the Eddington tensor, $c\,\kappa_{\nu}\,\rho$ in terms of the opacity $\kappa_{\nu}$ and  gas density $\rho$ is akin to the CR scattering rate $\bar{\nu}$, $q_{\nu} = dq/d\nu \equiv 4\pi\,\phi_{q}\,J_{\nu}$ is defined such that for photon number and energy $q=(n_{\gamma},\,e_{\gamma})$ we have corresponding $\phi_{q}=(1/h\nu,\,1)$, and ${\bf F}_{q} \equiv \phi_{q}\,c\,\int d\Omega\,{\bf n}\,I_{\nu} = c\,\langle {\bf n}\rangle_{\nu}\,q_{\nu}$ ($\langle {\bf n} \rangle_{\nu} \equiv (4\pi\,J_{\nu})^{-1}\,\int d\Omega\, {\bf n} \,I_{\nu}$) is the flux term. With these definitions, the spatial part of the first two moments of the non-relativistic radiation-MHD moments equations, as usually  written in the lab frame, are \citep{mihalas:1984oup..book.....M}:
\begin{align}
\label{eqn:rhd.q} \frac{\partial q_{\nu}}{\partial t} &= - \nabla\cdot {\bf F}^{q}_{\nu} + ... \\ 
\label{eqn:rhd.flux} \frac{\partial {\bf F}^{q}_{\nu} }{\partial t} &+ c^{2}\,\nabla \cdot (\mathbb{D}_{\nu}\,q_{\nu}) = -c\,\kappa_{\nu}\,\rho\,\left[  {\bf F}^{q}_{\nu} - q_{\nu}\,{\bf v}_{\rm gas} \cdot (\mathbb{I} + \mathbb{D}_{\nu}) \right] + ...
\end{align}
Note that the equations in the co-moving frame (to leading order in $\mathcal{O}({\bf v}_{\rm gas}/c)$) are equivalent to taking $\partial_{t} \rightarrow  D_{t}$ and dropping the ${\bf v}_{\rm gas}$ term above, so our discussion here applies equally to both cases. Eq.~\ref{eqn:rhd.q} is again just advection, and integrating over a frequency interval from $\nu^{-}$ to $\nu^{+}$, we immediately have $\partial_{t}q = -\nabla \cdot {\bf F}_{q}$ (with $q \equiv  \int_{\nu^{-}}^{\nu^{+}} d\nu\,q_{\nu}$, ${\bf F}_q \equiv  \int_{\nu^{-}}^{\nu^{+}} d\nu\,{\bf F}^q_{\nu}$), so we only need to consider Eq.~\ref{eqn:rhd.flux}.

\subsection{The Strong-Scattering and Flux-Limited Diffusion-Like Limit}
\label{sec:rad:flux.steady}

In the strong-scattering ``local steady-state'' limit for the flux,  we have the usual diffusive approximation with $\mathbb{D}_{\nu}\rightarrow \mathbb{I}/3$, ${\bf F}^{q}_{\nu}  \rightarrow (4/3)\,{\bf v}_{\rm gas}\,q_{\nu} - (c/3\,\kappa_{\nu}\,\rho) \nabla q_{\nu}$. Integrating this, we immediately obtain:
\begin{align}
{\bf F}_{q}^{i} &\rightarrow \frac{4}{3}\,{\bf v}_{\rm gas}^{i}\,q - \omega_{{\rm r},\,q}^{i}\frac{c}{3\,\kappa_{0}\,\rho}\,(\nabla q)^{i} \\ 
\label{eqn:omega.kappa.eff} \omega_{{\rm r},\,q}^{i} &\equiv \frac{\int_{\nu^{-}}^{\nu^{+}} d\nu\, \kappa_{\nu}^{-1}\,\phi_{q}\,(\nabla J_{\nu})^{i}}{\int_{\nu^{-}}^{\nu^{+}} d\nu\, \kappa_{0}^{-1}\,\phi_{q}\,(\nabla J_{\nu})^{i}} \\ 
\nonumber &\approx \frac{(1+\alpha_{J}+\alpha_{q}-\alpha_{\ell,i})}{(1+\alpha_{J}+\alpha_{q}-\alpha_{\ell,i}-\alpha_{\kappa})}
\,\frac{(\xi^{1+\alpha_{J}+\alpha_{q}-\alpha_{\ell,i}-\alpha_{\kappa}}-1)}{(\xi^{1+\alpha_{J}+\alpha_{q}-\alpha_{\ell,i}}-1)}\,\xi^{\alpha_{\kappa}/2} \\ 
\nonumber &\approx 1 - \frac{\alpha_{\kappa}}{12}\,\left(1+\alpha_{J}+\alpha_{q}-\alpha_{\ell,i} - \alpha_{\kappa}/2 \right)\,\left| \ln{\xi} \right|^{2}  + \mathcal{O}(|\ln{\xi}|^{4}) 
\end{align}
where $J_{\nu}\propto \nu^{\alpha_{J}}$, $\phi_{q}\propto \nu^{\alpha_{q}}$, $\kappa_{\nu} \propto \nu^{\alpha_{\kappa}}$, and $(\nabla J_{\nu})^{i} = J_{\nu}^{i}/\ell_{\nu}^{i}$ with $\ell_{\nu}^{i} \propto \nu^{\alpha_{\ell,i}}$ for each gradient component. 

We can in principle solve for each value of $\alpha_{\ell,i}$ as in \S~\ref{sec:local.steady.state.diffusion} above\footnote{Specifically computing $\ell_{q}^{i}$, $\ell_{q^{\prime}}^{i}$ and using
\begin{align}
\nonumber \frac{\ell^{i}_{q^{\prime}}}{\ell^{i}_{q}} &\approx \frac{\alpha_{1Jq}\,(\alpha_{1Jq^{\prime}}-\alpha_{\ell,i})}{\alpha_{1Jq^{\prime}}\,(\alpha_{1Jq}-\alpha_{\ell,i})}\,\frac{(\xi^{\alpha_{1Jq^{\prime}}}-1)\,(\xi^{\alpha_{1Jq}-\alpha_{\ell,i}}-1)}{(\xi^{\alpha_{1Jq}}-1)\,(\xi^{\alpha_{1Jq^{\prime}}-\alpha_{\ell,i}}-1)} \\ 
 &\approx 1 - \frac{\alpha_{\ell,i}}{12}\,\left(\alpha_{q}-\alpha_{q^{\prime}}\right)\,\left| \ln{\xi} \right|^{2} + \mathcal{O}(|\ln{\xi}|^{4})
\end{align}
with $\alpha_{1Jq} \equiv 1 + \alpha_{J} + \alpha_{q}$.}; but if we assume that either the dependence of gradient scale length on wavelength in the bin is small ($\alpha_{\ell,i}\sim 0$) or just that the gradient {\em direction} does not strongly depend on wavelength across the bin ($\alpha_{\ell,x}\approx \alpha_{\ell,y}\approx \alpha_{\ell,z} \approx \alpha_{\ell}$), then we can write  this in terms of a scalar ``effective'' $\kappa$, 
\begin{align}
{\bf F}_{q} &\sim \frac{4}{3}\,{\bf v}_{\rm gas}\,q - \frac{c}{3\,\kappa^{\rm eff}\,\rho}\,\nabla q \\ 
\label{eqn:kappa.eff.rosseland} \frac{1}{\kappa_{\rm eff}} &\equiv \frac{\int_{\nu^{-}}^{\nu^{+}} d\nu\, \kappa_{\nu}^{-1}\,\phi_{q}\,|\nabla J_{\nu}|}{\int_{\nu^{-}}^{\nu^{+}} d\nu\,\phi_{q}\,|\nabla J_{\nu}|}
= \frac{\omega_{{\rm r},\,q}}{\kappa_{0}}
\end{align}
(i.e.\ just Eq.~\ref{eqn:omega.kappa.eff} with $\alpha_{\ell,i} \rightarrow \alpha_{\ell}$). 

Now if we assume $J_{\nu}$ is blackbody-like, so $|\nabla J_{\nu}| \rightarrow  |{\rm d}J_{\nu}/{\rm d}T|\,|\nabla T|$, and consider the equation for the radiation energy density $q=e_{\gamma}$ (so $\phi_{q}=1$), Eq.~\ref{eqn:kappa.eff.rosseland} becomes immediately recognizable as the usual Rosseland mean opacity (the $|\nabla T|$ term factors out, being independent of $\nu$). So essentially, we have just generalized this convention for (1) an arbitrary non-blackbody intensity, and (2) other conserved radiation  quantities such as $n_{\gamma}$ ($\phi_{q} = 1/h\nu$, $\alpha_{q}=-1$), needed if we wish to correctly evolve the radiation spectrum as a piecewise power-law with two  degrees of freedom.

\subsection{The Weak-Scattering and M1-Like Limits}

Now consider cases where one wishes to evolve the flux Eq.~\ref{eqn:rhd.flux} explicitly, in e.g.\ first-moment (M1) or variable Eddington tensor (VET) or other related moments-based methods. Integrating, in component form, we can write:
\begin{align}
\frac{1}{c^{2}}\frac{\partial {\bf F}_{q}^{i} }{\partial t} &= - \left[ \nabla \cdot \left( \int_{\nu^{-}}^{\nu^{+}} d\nu\,\phi_{q,\nu}\,J_{\nu}\,\mathbb{D}_{\nu} \right) \right]^{i} \\
\nonumber & \  \ \ \ \ - \left[ \int_{\nu^{-}}^{\nu^{+}} d\nu\,\kappa_{\nu}\,\rho\,\phi_{q,\nu}\,J_{\nu}\,
\left\{ 
\langle {\bf n} \rangle_{\nu} - \frac{{\bf v}_{\rm gas} }{c} \cdot [ \mathbb{I} + \mathbb{D}_{\nu} ] 
\right\}
\right]^{i} \\ 
\nonumber &\equiv - \omega_{{\rm r},\nabla,q}^{i}\,\left[ \nabla  \cdot (\mathbb{D}_{0}\,q) \right]^{i} \\
\nonumber &\ \ \ \ \   - \frac{\kappa_{0}\,\rho}{c}\,\left[ \omega_{{\rm r},1,q}^{i}\,{\bf F}_{q}^{i} - \omega_{{\rm r},0,q}^{i}\,q_{0}\,{\bf v}_{\rm gas} \cdot (\mathbb{I}+\mathbb{D}_{0}) \right]
\end{align}
If we assume $\langle {\bf n} \rangle_{\nu}^{i} \propto \nu^{\alpha_{n,i}}$ (analogous to $\alpha_{\mu}$ for CRs), then we can write:
\begin{align}
\omega_{{\rm r},1,q}^{i} & \equiv  \frac{\int_{\nu^{-}}^{\nu^{+}} d\nu\, \kappa_{\nu}\,{\bf F}_{\nu}^{q,i}}{\int_{\nu^{-}}^{\nu^{+}} d\nu\, \kappa_{0}\,{\bf F}_{\nu}^{q,i}} \\ 
\nonumber &\approx \frac{(1+\alpha_{J}+\alpha_{q}+\alpha_{n,i})}{(1+\alpha_{J}+\alpha_{q}+\alpha_{n,i}+\alpha_{\kappa})}
\,\frac{(\xi^{1+\alpha_{J}+\alpha_{q}+\alpha_{n,i}+\alpha_{\kappa}}-1)}{(\xi^{1+\alpha_{J}+\alpha_{q}+\alpha_{n,i}}-1)}\,\xi^{-\alpha_{\kappa}/2} \\ 
\nonumber &\approx 1 + \frac{\alpha_{\kappa}}{12}\,\left(1+\alpha_{J}+\alpha_{q}+\alpha_{n,i} + \alpha_{\kappa}/2 \right)\,\left| \ln{\xi} \right|^{2}  + \mathcal{O}(|\ln{\xi}|^{4}) 
\end{align}
and we have an analogous consistency relation which determines $\alpha_{n,i}$ for each component of ${\bf F}_{q}^{i}$:
\begin{align}
\Psi_{qq^{\prime}}^{i} &\equiv {\left(\frac{{\bf F}^{i}_{q}}{q\,v_{0}} \right)}{\Big/}{\left(\frac{{\bf F}^{i}_{q^{\prime}}}{q^{\prime}\,v_{0}} \right)} = \frac{q^{\prime}\,{\bf F}^{i}_{q}}{q\,{\bf F}_{q^{\prime}}^{i}} \\ 
\nonumber & \equiv \frac{\left( \int_{\nu^{-}}^{\nu^{+}}\,d\nu\,\langle {\bf n} \rangle_{\nu}^{i}\,\phi_{q}\,J_{\nu} \right)}{\left( \int_{\nu^{-}}^{\nu^{+}}\,d\nu\,\langle {\bf n} \rangle_{\nu}^{i}\,\phi_{q^{\prime}}\,J_{\nu} \right)}\,
\frac{\left( \int_{\nu^{-}}^{\nu^{+}}\,d\nu\,\phi_{q^{\prime}}\,J_{\nu} \right)}{\left( \int_{\nu^{-}}^{\nu^{+}}\,d\nu\,\phi_{q}\,J_{\nu} \right)} \\
\nonumber &\approx 
\frac{(3+\alpha_{J}+\alpha_{q})}{(3+\alpha_{J}+\alpha_{q^{\prime}})}\,
\frac{(3+\alpha_{J}+\alpha_{q^{\prime}}+\alpha_{n,\,i})}{(3+\alpha_{J}+\alpha_{q}+\alpha_{n,\,i})}\, \\ 
\nonumber & \ \ \ \ \ \times \ 
\frac{(\xi^{3+\alpha_{J}+\alpha_{q^{\prime}}}-1)}{(\xi^{3+\alpha_{J}+\alpha_{q}}-1)}
\frac{(\xi^{3+\alpha_{J}+\alpha_{q}+\alpha_{n,\,i}}-1)}{(\xi^{3+\alpha_{J}+\alpha_{q^{\prime}}+\alpha_{n,\,i}}-1)}  \\
\nonumber &\approx 1 + \frac{\alpha_{n,\,i}}{12}\,\left(\alpha_{q}-\alpha_{q^{\prime}}\right)\,\left| \ln{\xi} \right|^{2} + \mathcal{O}(|\ln{\xi}|^{4})
\end{align}
If the fluxes ${\bf F}_{q}$ are explicitly evolved, we can then use $\Psi_{qq^{\prime}}^{i}$ to determine $\alpha_{n,i}$ and thus $\omega_{{\rm r},1,q}^{i}$, just as in our ``exact'' and ``integrated, approximate'' methods from \S~\ref{sec:solution.methods} above. If instead we wish to replace $\omega_{{\rm r},1,q}^{i}$ with its ``local flux steady-state'' value we see from \S~\ref{sec:rad:flux.steady} we would have $\alpha_{n,i} \rightarrow -(\alpha_{\ell,i}+\alpha_{\kappa})$ in  $\omega_{{\rm r},1,q}^{i}$.

The real challenge arises with the treatment of the Eddington tensor ($\mathbb{D}_{\nu}$) terms in $\omega_{{\rm r},\nabla,q}^{i}$ and $\omega_{{\rm r},0,q}^{i}\,q_{0}$. For CRs, it is worth emphasizing that the relation we wrote in \S~\ref{sec:setup},  $\mathbb{D} = \chi\,\mathbb{I} + (1-3\,\chi)\,\bhat\otimes\bhat$ is not some approximate closure: it is the {\em most general possible} form of $\mathbb{D}$ for a gyrotropic DF, and depends on a single scalar degree of freedom $\langle \mu^{2} \rangle$ (and likewise, its parallel gradient $\bhat\cdot\nabla\cdot\mathbb{D}$ introduces only a single scalar degree of freedom). Moreover, gyrotropy means that even for an arbitrarily anisotropic CR DF, $\mathbb{D} \propto \bhat\otimes \bhat$ depends on the magnetic field direction $\bhat$, which is of course CR-momentum-independent. On the other hand, for radiation, $\mathbb{D}$ has, in general, five independent degrees of freedom, and the $\nabla \cdot (\mathbb{D}_{\nu}\,q_{\nu})$ term introduces $\sim 10$ more.\footnote{These come from the dependence of $I_{\nu}$ on $\hat{n}$ (ray direction), and its (arbitrary) gradient. Since $\mathbb{D}$ is a symmetric $3x3$ matrix (in 3D) normalized to have trace unity (as it is defined by the moments of $I_{\nu}$) it has 5 degrees of freedom, and we have a similar number of degrees of freedom for each component of the vector gradient of $I_{\nu}$ which appears in $\nabla \cdot (\mathbb{D}_{\nu}\,q_{\nu})$.} So the problem is rather severely under-constrained. Moreover, even in the simplest possible highly-anisotropic case, where the radiation at a given $\nu$ is perfectly coherent (free-streaming in a single direction), we have $\mathbb{D}_{\nu} \sim \langle {\bf n} \rangle_{\nu} \otimes \langle {\bf n} \rangle_{\nu}$. But this (unlike $\bhat\otimes \bhat$) depends on an evolved property of the radiation flux itself ($\langle {\bf n} \rangle_{\nu}$), so it can depend on $\nu$, which means that we have no formal justification to neglect the variation in $\mathbb{D}_{\nu}$ across the bin. 

This is not a new problem: defining  a robust ``closure'' for $\mathbb{D}_{\nu}$ is arguably the central challenge for moments-based radiation or neutrino-hydrodynamics schemes \citep[see e.g.][]{wilson:1975.m1.closure,levermore:1984.FLD.M1,gnedin.abel.2001:otvet,rosdahl:m1.method.ramses,murchikova:m1.neutrino.transport.closures,foucart:2018.pic.closure}. And many of the most popular numerical methods use highly-approximate closures which only approach the exact solutions in very specific regimes (e.g.\ when $I_{\nu}$ is nearly-isotropic, or the radiation is a perfectly-coherent one-dimensional beam, etc.). So it is not clear if, in practice, we {\em could} solve for the correct $\omega_{{\rm r},\nabla,q}^{i}$ and $\omega_{{\rm r},0,q}^{i}$, even if we specified exactly some simple functional closure relation for $\mathbb{D}_{\nu}$. Thus, lacking another way to make progress, we will briefly consider -- without justification, we stress -- what we {\em would} obtain if we neglect the variations in $\mathbb{D}_{\nu}$ across each bin.

For the $\omega_{{\rm r},0,q}^{i}$ term, if we neglect variation  in $\mathbb{D}_{\nu}$ across the bin we can calculate it directly as 
$\omega_{{\rm r},0,q}^{i} \approx [(1+\alpha_{J}+\alpha_{q})/(1+\alpha_{J}+\alpha_{q}+\alpha_{\kappa})]\,[(\xi^{1+\alpha_{J}+\alpha_{q}+\alpha_{\kappa}}-1)/(\xi^{1+\alpha_{J}+\alpha_{q}}-1)]\,\xi^{-\alpha_{\kappa}/2} \approx 1 + (\alpha_{\kappa}/12)\,(1+\alpha_{J}+\alpha_{q} + \alpha_{\kappa}/2)\,|\ln{\xi}|^{2} + \mathcal{O}(|\ln\xi|^{4})$. But at this level of approximation, we can also just as well take $\omega_{{\rm r},0,q}^{i}\rightarrow \omega_{{\rm r},1,q}^{i}$, for the simple reason that for a non-relativistic ${\bf v}_{\rm gas}$ (the valid limit of our expressions), the ``advection'' term in ${\bf v}_{\rm gas}$ is only ever important in the strong-scattering, tightly-coupled regime, where we can (quite  accurately) assume the ``local steady-state'' approximation from above and, just like with CRs, this term reduces exactly to its ``bin-centered'' version, with $\omega_{{\rm r},0,q}^{i}/\omega_{{\rm r},1,q}^{i} \rightarrow 1$. 

The $\nabla \cdot (\mathbb{D}_{\nu}\,q_{\nu})$ term becomes trivial with $\omega_{{\rm r},\nabla,q}^{i}\rightarrow 1$ if we neglect variations in $\mathbb{D}_{\nu}$ across the bin. But we caution that while simple, this is much less ``safe'' an assumption than neglecting the variations for $\omega_{{\rm r},0,q}^{i}$. That is because this term is the dominant term controlling $\partial_{t} {\bf F}_{q}$ in the weak-scattering regime, which is  precisely where we said earlier it is {\em not} always safe to neglect variations in $\mathbb{D}_{\nu}$ with $\nu$. But for many moments-based methods, that regime is also where $\mathbb{D}_{\nu}$ is estimated rather poorly.  So this may not be a significant source of error relative to those pre-existing errors for methods like M1, but that remains to be tested.

\section{Conclusions}

We derive and test a simple improvement to numerical methods which dynamically evolve the CR spectrum, representing it as a piecewise power-law across momentum-space with standard advection/diffusion behavior in coordinate-space. Previous attempts to do so generally allow for smooth and exact evolution of the piecewise-power-law slopes under momentum-space operations (e.g.\ continuous and catastrophic losses, injection, etc.), but for the spatial terms adopted the ``bin-centered'' approximation which leads to errors in the local spectral shape when CR diffusion is important (or these methods sacrificed conservation or consistency with the underlying flux equations). We show that these errors are formally second-order in momentum-space, but they can be eliminated, allowing for smooth evolution of the CR spectra under diffusion, maintaining consistency with the underlying Vlasov equations and manifest conservation of CR number and energy (and current and momentum, in two-moment methods). 

The modification amounts to a set of three simple, scalar correction factors which, once computed, can be immediately applied (as e.g.\ a correction to the ``effective'' bin-centered diffusion coefficient, or to the scalar quantities whose gradients are calculated), which can be computed exactly entirely as a function  of actual evolved quantities in-code (i.e.\ there is no need to invoke new assumptions, or to implicitly evolve or take gradients of a ``finer grained'' distribution function). They require no fundamental modification to the numerical method adopted (and have no effect on its stability properties). 

The important conceptual addition is that the definitions of the conserved quantities and structure of the underlying equations for the DF impose a consistency requirement for how the mean pitch-angle $\langle \mu \rangle$ must vary across the bin, which allows us to derive these correction factors. We consider both exact formulations  of this constraint, and even simpler, approximate versions which still maintain manifest conservation and ensure consistency in all relevant limits when the CR flux equations are in local steady-state (e.g.\ on time/spatial scales larger than the CR scattering time/mean-free-path). We test these in a simple idealized problem and show they recover the desired behaviors, with negligible difference in computational expense.  All of the above applies both to one-moment methods which evolve a single scalar diffusion+streaming/advection equation (or Fokker-Planck-type equation) or two-moment methods which explicitly evolve the CR flux. 

We also extend this idea to similar methods which evolve radiation or neutrino hydrodynamics (again treating the spectrum as a piecewise power-law, attempting to simultaneously conserve both photon number and energy). We show that in the ``local flux steady-state'' or ``single-moment'' limit (aka the advective-diffusive limit for radiation transport), in which the intensity is close-to-isotropic, the appropriate correction terms can be derived and represent a generalization of the usual Rosseland mean opacity to arbitrary non-thermal spectra and other conserved quantities (e.g.\ photon number). However in the weak-scattering limit, the usual ambiguity in the form of the Eddington tensor makes the problem under-determined. The key difference is that we can safely assume CRs have a close-to-gyrotropic DF with respect to the magnetic field direction (which is, of course, CR momentum-independent) -- but there is no analogous constraint for radiation.

\acknowledgments{Support for PFH was provided by NSF Research Grants 1911233 \&\ 20009234, NSF CAREER grant 1455342, NASA grants 80NSSC18K0562, HST-AR-15800.001-A. Numerical calculations were run on the Caltech compute cluster ``Wheeler,'' allocations FTA-Hopkins supported by the NSF and TACC, and NASA HEC SMD-16-7592. Support for JS  was provided by Rutherford Discovery Fellowship RDF-U001804 and Marsden Fund grant UOO1727, which are managed through the Royal Society Te Ap\=arangi.}

\datastatement{The data supporting this article are available on reasonable request to the corresponding author.} 

\bibliography{ms_extracted}

\end{document}